# Nonlinear Hall quantum oscillations to probe topological Brown-Zak fermions in graphene moiré systems

Jinrui Zhong[1], Huimin Peng[1], Yuqing Hu[1], Qi Feng[1], Qiuli Li[1], Shihao Zhang[2], Qinsheng Wang[1], Jinhai Mao[3], Junxi Duan[1,✉], Yugui Yao[1,4,✉]

[1]Key Laboratory of Advanced Optoelectronic Quantum Architecture and Measurement (MOE), School of Physics, Beijing Institute of Technology, Beijing 100086, China

[2]School of Physics and Electronics, Hunan University, Changsha 410082, China

[3]School of Physical Sciences and CAS Center for Excellence in Topological Quantum Computation, University of Chinese Academy of Sciences, Beijing 100049, China

[4]International Center for Quantum Materials, Beijing Institute of Technology, Zhuhai, 519000, China

✉ Corresponding authors

E-mails: junxi.duan@bit.edu.cn;

ygyao@bit.edu.cn

## Abstract

Due to the deep connection with the quantum geometry of electronic Bloch wavefunctions, the second-order nonlinear Hall effect (NLHE) has been an attractive topic since its proposal. However, studies on NLHE under a magnetic field have been lacking. Given that quantum oscillations in the linear response regime have been proven to be useful tools in investigating electronic systems, searching for quantum oscillations in NLHE is of great interest and is expected to provide new avenues to unveil rich quantum geometric properties of novel quasiparticles. Here, we propose a new type of NLHE quantum oscillations and experimentally probe it in graphene moiré systems. It stems from the alternation of the dominant NLHE mechanisms with recurring Bloch states under magnetic field, which enables sensitive detection of Brown-Zak fermions, giving an onset field as low as 0.5 T. Most importantly, when the commensurability condition is satisfied, the nonlinear transport of Brown-Zak fermions is mainly governed by quantum geometric contributions. Our findings not only establish a new type of quantum oscillations, but also demonstrate the first experimental detection of the topological nature of Brown-Zak fermions, shedding light on the exploration of novel topological quasiparticles.

Due to its characterization of the electrons' motion by the geometry of their Bloch wavefunctions in Hilbert space, the quantum geometry, which is composed of Berry curvature and quantum metric, lays the foundation for modern condensed matter physics [1,2]. By directly revealing the quantum geometric distributions in momentum space, the high-order nonlinear transport effects have opened up new and hot frontiers, playing a central role in quantum geometry studies in recent years [3,4]. A representative one is the second-order nonlinear Hall effect (NLHE) [4-23]. This ac electric field triggered Hall response with a doubled response frequency evoked by the dipole moments of quantum geometry or scatterings has been demonstrated to be useful in topological-transition sensing and signal rectification [3,14,15,24]. However, previous literature of NLHE are generally conducted without magnetic field. Exploration of NLHE under a magnetic field, especially searching for new types of quantum oscillations in NLHE, is of great interest and importance but has been lacking. This is because, as fundamental quantum effects, quantum oscillations in the linear-response regime, e.g. Shubnikov-de Hass oscillations, have been proven to be of great significance in investigating electronic systems, such as mapping Fermi surface and detecting exotic isospin states and quasiparticles [25,26]. In light of the connection between NLHE and quantum geometry, quantum oscillations in NLHE are expected to provide new avenues to unveil rich quantum geometric properties of novel quasiparticles within the magnetic Brillouin zone.

Recently, Brown-Zak fermion, a quasiparticle in the Hofstadter spectrum, has attracted intense interest [27-36]. In general, introducing a magnetic field to an electronic system breaks the lattice translation symmetry, along with the collapse of Bloch theorem. However, under the commensurability condition where the magnetic flux per unit cell is commensurate with the flux quantum, the translation symmetry restores together with recurring Bloch states, forming many mini-band dispersions mimicking the one at zero magnetic fields, but with greatly folded magnetic Brillouin zone [37-39]. The quasiparticles, to describe the collective behaviors of electrons at commensurability conditions, are the so-called Brown-Zak fermions [29,30]. Consequently, they travel in straight trajectories or even ballistically over a long distance. Meanwhile, they inherit the quantum geometric properties of the electrons under zero field. Such intriguing characters make them a good candidate for developing novel electric devices under magnetic field. On experiments, the detection and modulation of Brown-Zak fermions is not feasible until the discovery of moiré superlattices with greatly enlarged unit cell. Previous literature have demonstrated Brown-Zak fermions' transport through the significant Brown-Zak oscillations (BZO) in linear response regime [27,28]. However, to reduce the overlap of minibands, a relatively high magnetic field is required, which inevitably induces the coexistence of Landau levels. Moreover, due to the long-distance-straight-trajectory transport of Brown-Zak fermions, their quantum geometric property is too weak to be uncovered.

In this paper, we propose a new type of NLHE quantum oscillations based on the dramatic alternation of the dominant NLHE mechanisms with and without recurring Bloch states, and we experimentally probe it in graphene moiré systems. With it, we have detected Brown-Zak fermions under a relatively low magnetic field, $\sim 0.5\ T$, much smaller than the general value ($B > 1.5\ T$) in the linear response regime [27,28,31-36]. Most importantly, through scaling analysis, we demonstrate that the NLHE of Brown-Zak fermions, i.e. under the commensurability condition, is dominated by quantum geometric contributions. Our results not only present a new type of quantum oscillations in the NLHE, but also provide the first experimental verification of the quantum geometric properties of Brown-Zak fermions, which shed light on further study of the Brown-Zak fermions and exploration of novel topological quasiparticles in twisted two-dimensional systems.

Unlike the formation of Landau levels, there is no gap-opening when Brown-Zak fermions emerge [27]. However, the quasiparticles in the system and their transport behavior change dramatically. When the lattice translation symmetry is broken by a general magnetic field, it is the electrons moving in the channel just as described by the semiclassical Drude model, and the transport properties are determined by scattering mechanisms. Once the commensurability condition is satisfied, Brown-Zak fermions emerge and dominate the transport phenomena. Since they can transport ballistically, scattering mechanisms play an insignificant role in the transport properties. In contrast, the multiple Brown-Zak minibands in the folded magnetic Brillouin zone strongly affect the quantum geometric properties which can be sensitively detected by the NLHE. Consequently, the alterations of the dominant mechanisms of NLHE with varying magnetic field can give rise to Brown-Zak type quantum oscillations in the NLHE, named the nonlinear Hall Brown-Zak oscillations (NHBZO), which is radically different from the effect from magnetic-field-induced conductance oscillations.

To detect Brown-Zak fermions and probe the NHBZO, we have fabricated large-angle twisted graphene devices, including twisted bilayer graphene (TBG) and twisted double bilayer graphene (TDBG). Twisted graphene consists of two single or few-layer graphene sheets stacked with an interlayer twisted angle, shaping a large-scale moiré superlattice, thus triggering exotic phenomenon like unconventional superconductivity, quantum anomalous Hall effect and Chern insulators [7,40-43]. Meanwhile, the NLHE based on Berry curvature dipole (BCD) or skew-scatterings has been widely observed at zero magnetic field in twisted graphene systems [4,7,8,15-19]. To minimize the disturbance from the sophisticated electronic correlation phenomenon emerging in small-angle twisted graphene systems, we deliberately make the twist angle of our devices to be larger than $2.5^{\circ}$. We will mainly focus on the results from the TDBG device in the main text. Figure 1(a) shows the linear longitudinal resistance diagram as a functions of carrier density $n$ and interlayer displacement fields $D$, where $n$ modulates the Fermi surface and $D$ modifies the Bloch band geometry. The gap opening with increasing $D$ at charge neutral point (CNP), together with the absence of superlattice band gaps within the experimental tuning range of $n$, prove that the sample is a TDBG with twist angle larger than $2.5^{\circ}$ (a brief discussion about the twist angle can be seen in Supplemental Material (SM) [44]). The sample quality is guaranteed by (1) the Landau fan diagram in SM Fig.S2 [44]; (2) the less than $0.08 \times 10^{12} cm^{-2}$ peak width at half-height of CNP gap at extremely high $D = -0.7V/nm$ (SM Fig.S7 [44]).

Next, we perform the second-order nonlinear transport measurement utilizing the second-harmonic technique [10-13]. An ac current $I^{\omega}$ with frequency $\omega$ is applied to the devices, and the lock-in amplifiers are used to simultaneously record the linear voltage $V^{\omega}$ and the nonlinear voltage $V^{2\omega}$ at doubled frequency. As a second-order nonlinear transport, the voltage $V^{2\omega}$ should have a parabolic dependence on the driving current, just as shown by the I-V curve at Fig.1(b). Meanwhile, under opposite current directions, $V^{2\omega}$ does not change sign. These features (and other confirmation in SM [44]) guarantee the second-order nonlinear transport nature in our measurements.

Then, we perform magneto-transport measurements at fixed $n$ and $D$, plotted in Fig.1(c). Two points should be noted. (1) The second-order longitudinal voltage $V_x^{2\omega}$ (inset in Fig.1(c)) is completely absent. It remains absent or extremely weak during all our measurements (see

SM[44]). These only-in-transverse-direction responses indicate the Hall effect nature of $V_y^{2\omega}$, i.e. a NLHE. (2) At low magnetic fields, the linear Hall resistivity $\rho_{xy}$ (blue line) is dominated by the ordinary Hall effect. On the contrary, the $V_y^{2\omega}$ (red line) shows an extremely low fields oscillation behavior, with its onset well below 0.5 T. Meanwhile, the $V_y^{2\omega}$ maintains its sign when varying the magnetic field directions. In addition, the normalized response, $V_y^{2\omega}/(I^{\omega})^2$, shows no dependence on the magnitude and frequency of $I^{\omega}$ (see SM Fig.S3 [44]). The oscillations are well-defined even under $I^{\omega} = 50$ nA, which is an extremely low current for nonlinear transport experiments. Therefore, the oscillations in $V_y^{2\omega}$ are one kind of magnetic-field-induced oscillations in the NLHE. Below, for convenience, the observed oscillations in $V_y^{2\omega}$ is denoted as NHO.

We now carefully examine this NHO. The periodicity in 1/B, which is shown in Fig.1(d) by the differential voltage $dV_y^{2\omega}/dB$ curve with 1/B as horizontal coordinate, reveals its quantum oscillation characteristics. Within the Hofstadter's spectrum, quantum oscillations with $1/B$ periodicity in the linear response regime can have two mechanisms in principle: (1) gapped states following the Streda formula, whose $B$ trajectory changes with different $n$, such as Landau levels and Chern insulators; (2) the recurrence of Brown-Zak fermions, i.e. the Brown-Zak oscillations (BZO) [27,28], whose $B$ trajectory is independent on $n$, because the Brown-Zak fermions emerge only under the commensurability condition. Therefore, we plot $V_y^{2\omega}$ as a function of $n$ and $1/B$ in Fig.2(a). The differential diagram $dV_y^{2\omega}/dB$ is also drawn in Fig.2(b) for better resolution. Obviously, the field-induced $V_y^{2\omega}$ oscillations have an $n$-independent periodicity, and the peaks appear only around fixed $n$-independent $B$ values. This $n$-independent periodicity strongly distinguishes the NHO from all the magneto-oscillatory effects arising from either Landau quantization or commensurability oscillation, but mimicking the BZO in the linear regime [27,28,31-36]. The electron-hole asymmetry that the oscillations are clearer for electrons is also consistent with the Brown-Zak fermions' behavior in the literature [27,28,31-36].

We now confirm the relationship between the NHO and Brown-Zak fermions. (1) The Brown-Zak fermions' transport is an orbital behavior, which depends only on the out-of-plane component of $B$. To exclude possible contributions such as isospin order and the nonlinear planar Hall effects [45,46], we measure the NHO at fixed $n$ and $D$ with different angles $\theta$ between $B$ and the normal direction of the device plane. In Fig.3(a), we plot $dV_y^{2\omega}/dB$ as a function of $1/B$ with different $\theta$. When $\theta$ is increased, the NHO experiences a 'shrinkage', shifting to higher $B$. Through fast Fourier transform (FFT), we can extract the frequency, $B_\theta$, of these oscillations. As presented in Fig.3(b), $B_\theta$ showing a perfect $B_0 = B_\theta \cos\theta$ relation, where $B_0$ is the frequency at $\theta = 0^\circ$. The NHO is completely absent under in-plane $B$ (Fig.S6 [44]). These confirm that the NHO is a purely orbital effect. (2) The Brown-Zak fermions' transport is not a Fermi surface behavior. Its trajectory merely depends on the size of unit cell and the magnetic field values, but not on the Bloch band geometry. Therefore, we plot the $dV_y^{2\omega}/dB$ under different displacement field $D$ in Fig.3(c). In TDBG, $D$ can effectively alter the band dispersion. However, for the NHO, the oscillation trajectories remain the same even at high $D$. This $D$-independence also eliminates the trivial contributions from poor ohmic contacts and lattice potential from hBN substrates.

With these evidences, the NHO illustrate nothing but the second-order nonlinear Hall quantum oscillations from Brown-Zak fermions, thus, the NHBZO. We should note the difference between the NHBZO and the BZO. As shown in Fig. 2(c), the linear longitudinal resistance $R_{xx}$, as well as the linear Hall resistance $\rho_{xy}$ (Fig. S4 [44]), at low fields shows no detectable oscillations. The onset field of the NHBZO is lower than $0.5\ T$, much smaller than the general one ($B > 1.5\ T$) of the BZO in the literature [27,28,31-36]. It reveals that NHBZO is a more sensitive method to detect the transport properties of Brown-Zak fermions. This radical difference also indicates that the NHBZO cannot be attributed to the effect from conductivity oscillations in the BZO, although NLHE depends on the conductivity. We need to mention that the NHBZO has also been observed in TBG device, where it coexists with quantum oscillations originated from Landau levels in second-order response (Fig. S5 [44]).

Finally, we illustrate that the observed NHBZO comes from the expected mechanism. According to theoretical studies, the second-order nonlinear transverse transport has various mechanisms, including BCD, quantum metric dipole (QMD), skew-scattering, side jump, nonlinear Drude, and so on [4,47,48], which can be distinguished regarding to their power-law scaling on the scattering time $\tau$ [4,5,47,49-53]. The analysis is conventionally conducted by changing the device temperature to tune $\tau$ [11-18]. Due to the Hall nature of the NHBZO, contributions from skew-scattering and side-jump mechanisms can firstly be ignored (see more detailed discussion in SM[44]). As a result, for a second-order nonlinear Hall transport at low temperatures ($T \leq 10$ K in our case that the dynamical scattering, e.g. phonons, can be neglected) under magnetic field, the scaling relation of $\chi_{yxx}^{2\omega}$ has the form [4,47]:

$$\chi_{yxx}^{2\omega} = -\sigma \frac{V_y^{2\omega} L^2}{(V_x)^2 W} = a\sigma^2 + b\sigma + c \quad (1)$$

where $\sigma$ is the linear longitudinal conductivity, and $L$ and $W$ the length and width of the Hall bar, respectively. Here, considering that $\sigma \sim ne^2\tau/m$, $\tau$ has been replaced by $\sigma$ when the carrier density $n$ is fixed. The coefficient $a$ represents the nonlinear Drude contribution, while $b$ the BCD and $c$ the QMD. Since $\sigma$ shows no oscillations at low fields, the NHBZO should come from the changes of the coefficients.

To identify these changes, we have simultaneously measured $\chi_{yxx}^{2\omega}$ and $\sigma$ as functions of $n$ and $B$ at different temperatures from $T = 1.6$ K to 10 K. The blue circles in Fig.4(a) show the $\chi_{yxx}^{2\omega}$ vs. $\sigma$ with $n = 1.5 \times 10^{12}\ cm^{-2}$ at $B = 0.74$ T where Brown-Zak fermions emerge. Clearly, $\chi_{yxx}^{2\omega}$ shows prominent linear dependence on $\sigma$ with a considerably nonzero intercept (inset in Fig. 4(a)), which is best fitted by a linear line (blue line), indicating quantum-geometric contributions dominance. Since the emergence of Brown-Zak fermions requires the commensurability condition, a slight deviation could destroy these quasiparticles and their topological transport. This is revealed when the magnetic field is tuned away from $B = 0.74$ T with a tiny $\Delta B = 0.04$ T, as shown by the red circles in Fig.4(a), that the scaling relation is no longer linear but becomes parabolic (the red line in Fig. 4(a)). It means that nonlinear Drude contribution becomes dominant, as is expected by general semiclassical transport theory [54-57]. More data points at several other fixed magnetic fields show consistent behavior, as shown in Fig.S16 [44].

To have a direct view of these contributions as functions of $n$ and $B$, we perform polynomial fitting $\chi_{yxx}^{2\omega} = a\sigma^2 + b\sigma + c$ at each $(n, B)$ point based on the measured $\chi_{yxx}^{2\omega}$ and $\sigma$ at different temperatures, and plot them in Fig.4(b)-(d), respectively. Here, since the three coefficients differ in units, for a better comparison, we employ the normalized value $c/\chi_{yxx}^{2\omega}$, $b\sigma/\chi_{yxx}^{2\omega}$, and $a\sigma^2/\chi_{yxx}^{2\omega}$ to show their weights in $\chi_{yxx}^{2\omega}$ (Fig.S17 [44]). The results show three interesting features. First, the three contributions show prominent strips parallel to the $n$ axis. They oscillate strongly along $B$ but vary smoothly along $n$. Second, $b\sigma/\chi_{yxx}^{2\omega}$ and $a\sigma^2/\chi_{yxx}^{2\omega}$ show opposite dependence on $B$. In the vicinity of the $B$ where Brown-Zak fermions emerge, $b\sigma/\chi_{yxx}^{2\omega}$ is giant but $a\sigma^2/\chi_{yxx}^{2\omega}$ even reaches zero. Once $B$ is tuned away from these regimes, $b\sigma/\chi_{yxx}^{2\omega}$ is strongly suppressed but $a\sigma^2/\chi_{yxx}^{2\omega}$ becomes dominant. Third, $c/\chi_{yxx}^{2\omega}$ and $b\sigma/\chi_{yxx}^{2\omega}$ behave similarly.

These three features undoubtedly confirm our proposed origination of the observed NHBZO. The existence of field-induced oscillations with a $n$-independent periodicity is unique for Brown-Zak fermions. Thus, the oscillations of these coefficients originate from the recurring of Brown-Zak fermions. Under general magnetic field (away from the commensurability condition), the dominance of the $a\sigma^2$ term indicates that electrons move in the channel just as described by the Drude model. Once the commensurability condition is met, Brown-Zak fermions emerge and the Bloch states recur. The absence of $a\sigma^2$ agrees with the reported results that Brown-Zak fermions transport in a reduced magnetic field and scattering effects are negligible [27]. On the contrary, the large weight of the $b\sigma$ and $c$ terms indicate that the quantum geometric properties, including BCD and QMD, play significant roles in the NLHE of Brown-Zak fermions, which gives the first experimental demonstration of the quantum geometric properties of Brown-Zak fermions. Therefore, it is the dramatic alternation of the quasiparticles and their transport behaviors, other than the change of conductivity, that generates the observed NHBZO. This mechanism is distinctly different from the one of the BZO, which makes the NHBZO a more sensitive tool to study Brown-Zak fermions (see Note 6 in SM for more discussions). More importantly, as it shows the generalizability, twist-angle-tolerance and Fermi-surface-insensitivity, we expect this intriguing NHBZO to be observed in other moiré systems, especially the twisted transition metal dichalcogenides which share a lot of similarities with twisted graphene families while having more broken symmetries.

To conclude, we have proposed and experimentally realized a new type of quantum oscillations in NLHE, termed NHBZO. It arises from the alternation of the quasiparticles and their transport behaviors. Such an intriguing mechanism makes NHBZO a more sensitive tool to investigate Brown-Zak fermions. Most importantly, when Brown-Zak fermions emerge, the NLHE is dominated by the quantum geometric contributions. This provides the first experimental detection of the quantum geometric properties of Brown-Zak fermions. Our findings not only expand the family of quantum oscillations, but also complement a missing piece of puzzles about quantum geometric behaviors within the Hofstadter's spectrum, which will inspire new research frontiers of NLHE such as further studies of the quantum geometric properties of Brown-Zak fermions and exploration of novel topological quasiparticles.

## Acknowledgement

We thank the helpful discussion with Yang Gao from University of Science and Technology of China, Xiao Li and Yingwen Zhang from City University of Hong Kong, Juncheng Li and Cong Chen from University of Hong Kong.

## References

[1] P. Torma, Essay: Where Can Quantum Geometry Lead Us?, Phys. Rev. Lett. **131**, 240001 (2023).
[2] T. Liu, X.-B. Qiang, H.-Z. Lu, and X. C. Xie, Quantum geometry in condensed matter, National Science Review **12**, nwae334 (2025).
[3] M. Suarez-Rodriguez, F. de Juan, I. Souza, M. Gobbi, F. Casanova, and L. E. Hueso, Nonlinear transport in non-centrosymmetric systems, Nat. Mater. **24**, 1005 (2025).
[4] Y. Jiang, T. Holder, and B. Yan, Revealing Quantum Geometry in Nonlinear Quantum Materials, Reports on Progress in Physics **88**, 076502 (2025).
[5] I. Sodemann and L. Fu, Quantum Nonlinear Hall Effect Induced by Berry Curvature Dipole in Time-Reversal Invariant Materials, Phys. Rev. Lett. **115**, 216806 (2015).
[6] Z. Z. Du, H.-Z. Lu, and X. C. Xie, Nonlinear Hall effects, Nat. Rev. Phys. **3**, 744 (2021).
[7] P. C. Adak, S. Sinha, A. Agarwal, and M. M. Deshmukh, Tunable moiré materials for probing Berry physics and topology, Nat. Rev. Mater. **9**, 481 (2024).
[8] L. Du, Z. Huang, J. Zhang, F. Ye, Q. Dai, H. Deng, G. Zhang, and Z. Sun, Nonlinear physics of moire superlattices, Nat. Mater. **23**, 1179 (2024).
[9] A. Bandyopadhyay, N. B. Joseph, and A. Narayan, Non-linear Hall effects: Mechanisms and materials, Materials Today Electronics **8**, 100101 (2024).
[10] Q. Ma, S. Y. Xu, H. Shen, D. MacNeill, V. Fatemi, T. R. Chang, A. M. Mier Valdivia, S. Wu, Z. Du, C. H. Hsu, S. Fang, Q. D. Gibson, K. Watanabe, T. Taniguchi, R. J. Cava, E. Kaxiras, H. Z. Lu, H. Lin, L. Fu, N. Gedik, and P. Jarillo-Herrero, Observation of the nonlinear Hall effect under time-reversal-symmetric conditions, Nature **565**, 337 (2019).
[11] K. Kang, T. Li, E. Sohn, J. Shan, and K. F. Mak, Nonlinear anomalous Hall effect in few-layer $WTe_2$, Nat. Mater. **18**, 324 (2019).
[12] A. Gao, Y.-F. Liu, J.-X. Qiu, B. Ghosh, T. V. Trevisan, Y. Onishi, C. Hu, T. Qian, H.-J. Tien, S.-W. Chen, M. Huang, D. Bérubé, H. Li, C. Tzschaschel, T. Dinh, Z. Sun, S.-C. Ho, S.-W. Lien, B. Singh, K. Watanabe, T. Taniguchi, D. C. Bell, H. Lin, T.-R. Chang, C. R. Du, A. Bansil, L. Fu, N. Ni, P. P. Orth, Q. Ma, and S.-Y. Xu, Quantum metric nonlinear Hall effect in a topological antiferromagnetic heterostructure, Science **381**, 181 (2023).
[13] N. Wang, D. Kaplan, Z. Zhang, T. Holder, N. Cao, A. Wang, X. Zhou, F. Zhou, Z. Jiang, C. Zhang, S. Ru, H. Cai, K. Watanabe, T. Taniguchi, B. Yan, and W. Gao, Quantum-metric-induced nonlinear transport in a topological antiferromagnet, Nature **621**, 487 (2023).
[14] D. Kumar, C. H. Hsu, R. Sharma, T. R. Chang, P. Yu, J. Wang, G. Eda, G. Liang, and H. Yang, Room-temperature nonlinear Hall effect and wireless radiofrequency rectification in Weyl semimetal $TaIrTe_4$, Nat. Nanotechnol. **16**, 421 (2021).
[15] S. Sinha, P. C. Adak, A. Chakraborty, K. Das, K. Debnath, L. D. V. Sangani, K. Watanabe, T. Taniguchi, U. V. Waghmare, A. Agarwal, and M. M. Deshmukh, Berry curvature dipole senses topological transition in a moiré superlattice, Nat. Phys. **18**, 765 (2022).
[16] J. Duan, Y. Jian, Y. Gao, H. Peng, J. Zhong, Q. Feng, J. Mao, and Y. Yao, Giant Second-Order Nonlinear Hall Effect in Twisted Bilayer Graphene, Phys. Rev. Lett. **129**, 186801 (2022).
[17] M. Huang, Z. Wu, X. Zhang, X. Feng, Z. Zhou, S. Wang, Y. Chen, C. Cheng, K. Sun, Z. Y. Meng, and N. Wang, Intrinsic Nonlinear Hall Effect and Gate-Switchable Berry Curvature Sliding in Twisted Bilayer Graphene, Phys. Rev. Lett. **131**, 066301 (2023).

[18] J. Zhong, S. Zhang, J. Duan, H. Peng, Q. Feng, Y. Hu, Q. Wang, J. Mao, J. Liu, and Y. Yao, Effective Manipulation of a Colossal Second-Order Transverse Response in an Electric-Field-Tunable Graphene Moire System, Nano. Lett. **24**, 5791 (2024).
[19] N. J. Zhang, J. X. Lin, D. V. Chichinadze, Y. Wang, K. Watanabe, T. Taniguchi, L. Fu, and J. I. A. Li, Angle-resolved transport non-reciprocity and spontaneous symmetry breaking in twisted trilayer graphene, Nat. Mater. **23**, 356 (2024).
[20] M. Huang, Z. Wu, J. Hu, X. Cai, E. Li, L. An, X. Feng, Z. Ye, N. Lin, K. T. Law, and N. Wang, Giant nonlinear Hall effect in twisted bilayer $WSe_2$, National Science Review **10**, nwac232 (2022).
[21] L. Min, H. Tan, Z. Xie, L. Miao, R. Zhang, S. H. Lee, V. Gopalan, C. X. Liu, N. Alem, B. Yan, and Z. Mao, Strong room-temperature bulk nonlinear Hall effect in a spin-valley locked Dirac material, Nat. Commun. **14**, 364 (2023).
[22] L. Min, Y. Zhang, Z. Xie, S. V. G. Ayyagari, L. Miao, Y. Onishi, S. H. Lee, Y. Wang, N. Alem, L. Fu, and Z. Mao, Colossal room-temperature non-reciprocal Hall effect, Nat. Mater. **23**, 1671 (2024).
[23] B. Cheng, Y. Gao, Z. Zheng, S. Chen, Z. Liu, L. Zhang, Q. Zhu, H. Li, L. Li, and C. Zeng, Giant nonlinear Hall and wireless rectification effects at room temperature in the elemental semiconductor tellurium, Nat. Commun. **15**, 5513 (2024).
[24] X. F. Lu, C. P. Zhang, N. Wang, D. Zhao, X. Zhou, W. Gao, X. H. Chen, K. T. Law, and K. P. Loh, Nonlinear transport and radio frequency rectification in BiTeBr at room temperature, Nat. Commun. **15**, 245 (2024).
[25] D. Shoenberg, *Magnetic oscillations in metals* (Cambridge university Press, Cambridge, 1984).
[26] M. Bocarsly, I. Roy, V. Bhardwaj, M. Uzan, P. Ledwith, G. Shavit, N. Banu, Y. Zhou, Y. Myasoedov, K. Watanabe, T. Taniguchi, Y. Oreg, D. E. Parker, Y. Ronen, and E. Zeldov, Coulomb interactions and migrating Dirac cones imaged by local quantum oscillations in twisted graphene, Nat. Phys. **21**, 421 (2025).
[27] R. Krishna Kumar, X. Chen, G. H. Auton, A. Mishchenko, D. A. Bandurin, S. V. Morozov, Y. Cao, E. Khestanova, M. Ben Shalom, A. V. Kretinin, K. S. Novoselov, L. Eaves, I. V. Grigorieva, L. A. Ponomarenko, V. I. Fal'ko, and A. K. Geim, High-temperature quantum oscillations caused by recurring Bloch states in graphene superlattices, Science **357**, 181 (2017).
[28] J. Barrier, P. Kumaravadivel, R. Krishna Kumar, L. A. Ponomarenko, N. Xin, M. Holwill, C. Mullan, M. Kim, R. V. Gorbachev, M. D. Thompson, J. R. Prance, T. Taniguchi, K. Watanabe, I. V. Grigorieva, K. S. Novoselov, A. Mishchenko, V. I. Fal'ko, A. K. Geim, and A. I. Berdyugin, Long-range ballistic transport of Brown-Zak fermions in graphene superlattices, Nat. Commun. **11**, 5756 (2020).
[29] X. Chen, J. R. Wallbank, A. A. Patel, M. Mucha-Kruczyński, E. McCann, and V. I. Fal'ko, Dirac edges of fractal magnetic minibands in graphene with hexagonal moiré superlattices, Phys. Rev. B **89**, 075401 (2014).
[30] T. Fabian, M. Kausel, L. Linhart, J. Burgdörfer, and F. Libisch, Half-integer Wannier diagram and Brown-Zak fermions of graphene on hexagonal boron nitride, Phys. Rev. B **106**, 165412 (2022).
[31] C. Mullan, S. Slizovskiy, J. Yin, Z. Wang, Q. Yang, S. Xu, Y. Yang, B. A. Piot, S. Hu, T. Taniguchi, K. Watanabe, K. S. Novoselov, A. K. Geim, V. I. Fal'ko, and A. Mishchenko, Mixing

of moiré-surface and bulk states in graphite, Nature **620**, 756 (2023).
[32] W. Shi, S. Kahn, N. Leconte, T. Taniguchi, K. Watanabe, M. Crommie, J. Jung, and A. Zettl, High-Order Fractal Quantum Oscillations in Graphene/BN Superlattices in the Extreme Doping Limit, Phys. Rev. Lett. **130**, 186204 (2023).
[33] M. K. Jat, P. Tiwari, R. Bajaj, I. Shitut, S. Mandal, K. Watanabe, T. Taniguchi, H. R. Krishnamurthy, M. Jain, and A. Bid, Higher order gaps in the renormalized band structure of doubly aligned hBN/bilayer graphene moiré superlattice, Nat. Commun. **15**, 2335 (2024).
[34] Y. Jeong, H. Park, T. Kim, K. Watanabe, T. Taniguchi, J. Jung, and J. Jang, Interplay of valley, layer and band topology towards interacting quantum phases in moiré bilayer graphene, Nat. Commun. **15**, 6351 (2024).
[35] H. Tian, E. Codecido, D. Mao, K. Zhang, S. Che, K. Watanabe, T. Taniguchi, D. Smirnov, E.-A. Kim, M. Bockrath, and C. N. Lau, Dominant 1/3-filling correlated insulator states and orbital geometric frustration in twisted bilayer graphene, Nat. Phys. **20**, 1407 (2024).
[36] Y. Ma, M. Huang, X. Zhang, W. Hu, Z. Zhou, K. Feng, W. Li, Y. Chen, C. Lou, W. Zhang, H. Ji, Y. Wang, Z. Wu, X. Cui, W. Yao, S. Yan, Z. Y. Meng, and N. Wang, Magnetic Bloch states at integer flux quanta induced by super-moiré potential in graphene aligned with twisted boron nitride, Nat. Commun. **16**, 1860 (2025).
[37] E. Brown, Bloch Electrons in a Uniform Magnetic Field, Phys. Rev. **133**, A1038 (1964).
[38] J. Zak, Magnetic Translation Group, Phys. Rev. **134**, A1602 (1964).
[39] D. R. Hofstadter, Energy levels and wave functions of Bloch electrons in rational and irrational magnetic fields, Phys. Rev. B **14**, 2239 (1976).
[40] E. Y. Andrei and A. H. MacDonald, Graphene bilayers with a twist, Nat. Mater. **19**, 1265 (2020).
[41] J. Liu and X. Dai, Orbital magnetic states in moiré graphene systems, Nat. Rev. Phys. **3**, 367 (2021).
[42] P. Törmä, S. Peotta, and B. A. Bernevig, Superconductivity, superfluidity and quantum geometry in twisted multilayer systems, Nat. Rev. Phys. **4**, 528 (2022).
[43] K. P. Nuckolls and A. Yazdani, A microscopic perspective on moiré materials, Nat. Rev. Mater. **9**, 460 (2024).
[44] See Supplemental Material at [URL] for details on device fabrication, electric transport measurement, components of nonlinear Hall conductivity, symmetry discussion about the existence of Berry curvature dipole, discussions about twist angles, exclude other extrinsic mechanisms, About other magneto-oscillatory phenomena in nonlinear transport, On the low onset field of NHBZO, and other supporting data images S1-S19.
[45] P. He, S. S. L. Zhang, D. Zhu, S. Shi, O. G. Heinonen, G. Vignale, and H. Yang, Nonlinear Planar Hall Effect, Phys. Rev. Lett. **123**, 016801 (2019).
[46] Y.-X. Huang, X. Feng, H. Wang, C. Xiao, and S. A. Yang, Intrinsic Nonlinear Planar Hall Effect, Phys. Rev. Lett. **130**, 126303 (2023).
[47] Z.-H. Gong, Z. Z. Du, H.-P. Sun, H.-Z. Lu, and X. C. Xie, Nonlinear transport theory at the order of quantum metric, arXiv:2410.04995
[48] X. B. Qiang, T. Liu, Z. X. Gao, H. Z. Lu, and X. C. Xie, A Clarification on Quantum-Metric-Induced Nonlinear Transport, Adv. Sci. **13**, e14818 (2025).
[49] Z. Z. Du, C. M. Wang, S. Li, H. Z. Lu, and X. C. Xie, Disorder-induced nonlinear Hall effect with time-reversal symmetry, Nat. Commun. **10**, 3047 (2019).

[50] C. Xiao, H. Zhou, and Q. Niu, Scaling parameters in anomalous and nonlinear Hall effects depend on temperature, Phys. Rev. B **100**, 161403(R) (2019).
[51] Z. Z. Du, C. M. Wang, H.-P. Sun, H.-Z. Lu, and X. C. Xie, Quantum theory of the nonlinear Hall effect, Nat. Commun. **12**, 5038 (2021).
[52] D. Kaplan, T. Holder, and B. Yan, Unification of Nonlinear Anomalous Hall Effect and Nonreciprocal Magnetoresistance in Metals by the Quantum Geometry, Phys. Rev. Lett. **132**, 026301 (2024).
[53] Y.-X. Huang, C. Xiao, S. A. Yang, and X. Li, Scaling law and extrinsic mechanisms for time-reversal-odd second-order nonlinear transport, Phys. Rev. B **111**, 155127 (2025).
[54] Y. Gao, S. A. Yang, and Q. Niu, Field induced positional shift of Bloch electrons and its dynamical implications, Phys. Rev. Lett. **112**, 166601 (2014).
[55] Y. Gao, Semiclassical dynamics and nonlinear charge current, Frontiers of Physics **14**, 33404 (2019).
[56] S. Nandy and I. Sodemann, Symmetry and quantum kinetics of the nonlinear Hall effect, Phys. Rev. B **100**, 195117 (2019).
[57] J. Jia, L. Xiang, Z. Qiao, and J. Wang, Equivalence of semiclassical and response theories for second-order nonlinear ac Hall effects, Phys. Rev. B **110**, 245406 (2024).

# Figures and captions

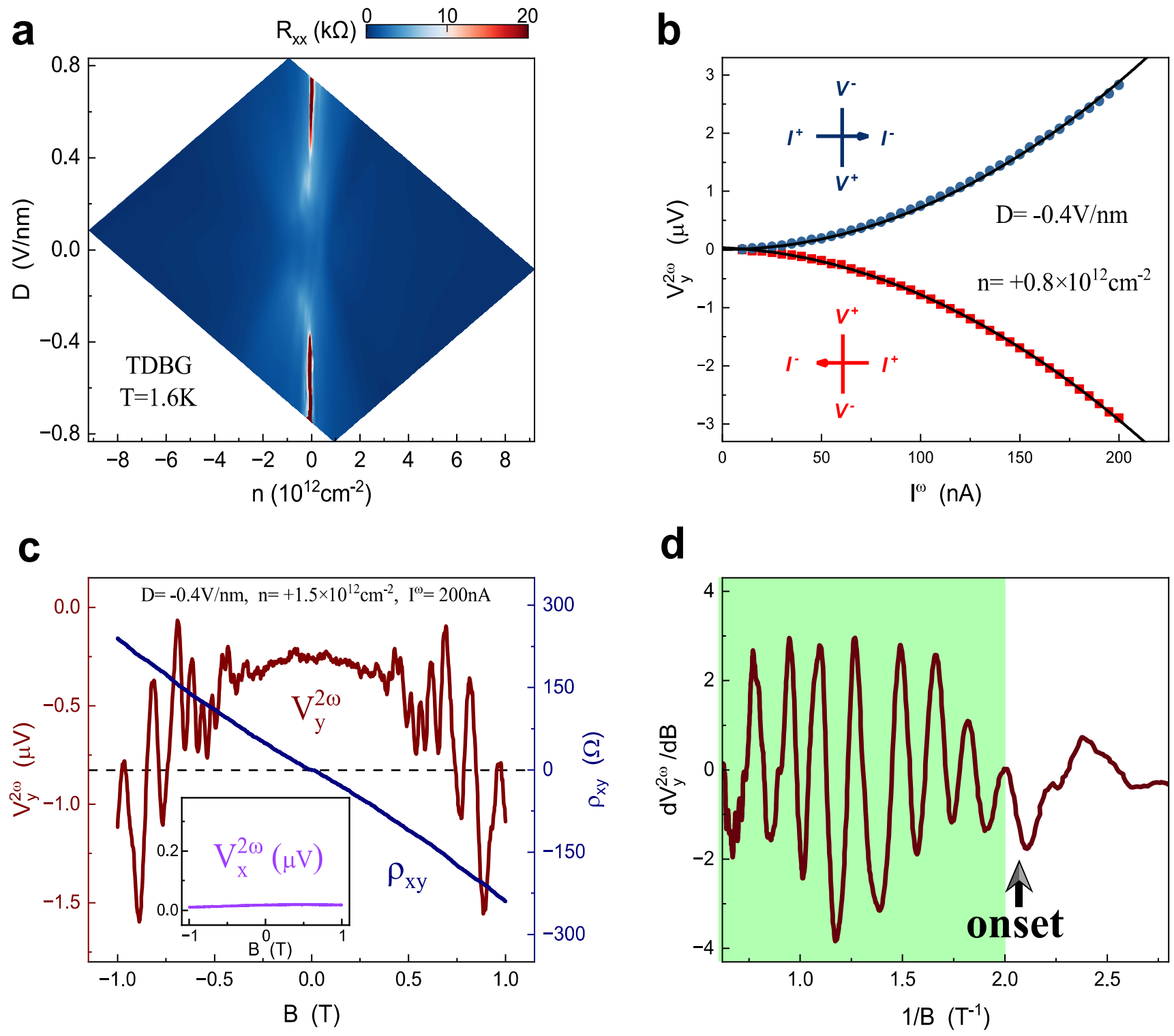

**FIG. 1. The nonlinear Hall oscillations in twisted double bilayer graphene. (a)** $R_{xx}$ as a function of carrier density $n$ and displacement fields $D$ in TDBG device at T=1.6K, B=0T. **(b)** $V_y^{2\omega}$ when sweeping current $I^{\omega}$. Blue and red data points are measured with opposite current direction and measurement geometry. The black lines are parabolic fitting lines. **(c)** $V_y^{2\omega}$ (red, left axis) and $\rho_{xy}$ (blue, right axis) as a function of $B$. Inset: $V_x^{2\omega}$ vs $B$. **(d)** $dV_y^{2\omega}/dB$ as a function of $1/B$. Green shadow denotes prominent oscillation region.

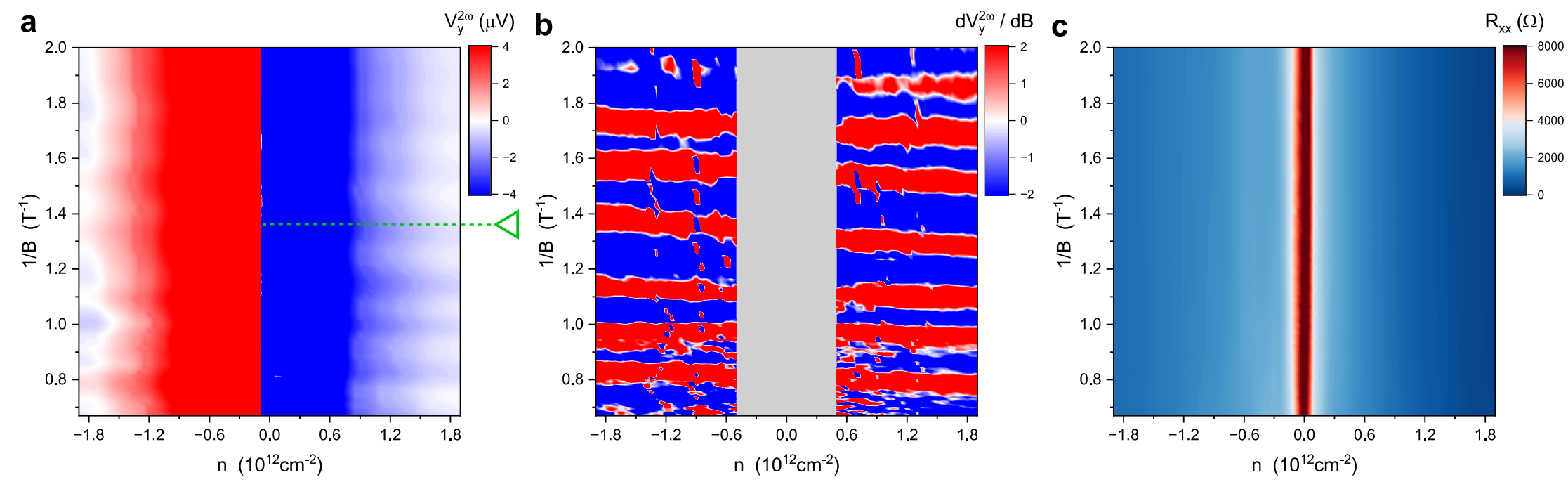

**FIG. 2. Nonlinear Hall Brown-Zak oscillation diagrams in TDBG. (a)** $V_y^{2\omega}$ as a function of $n$ and reciprocal of magnetic fields $1/B$ at $D = -0.4V/nm$ in TDBG. The driving currents are 200nA. The green dashed line and triangle indicate the magnetic field value $0.74T$ in Fig.4(a). (**b**) $dV_y^{2\omega}/dB$ diagram within the oscillation region calculated from (**a**). (**c**) diagram of $R_{xx}$ in TDBG at $D = -0.4V/nm$.

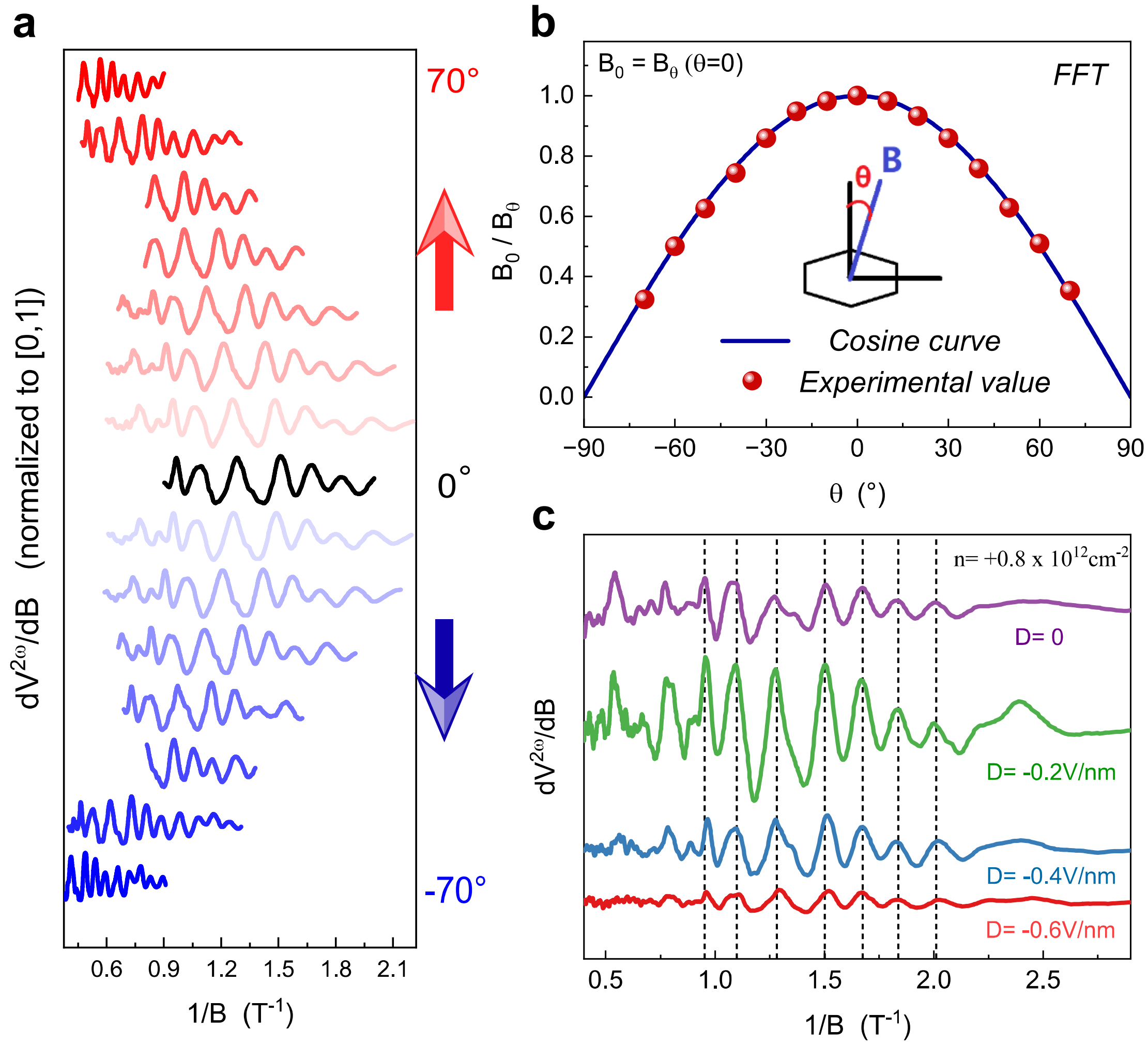

**FIG. 3. Independence of in-plane magnetic field component and displacement fields in TDBG. (a)** $dV_y^{2\omega}/dB$ as a function of $1/B$ when rotating the magnetic field direction from out-of-plane (black line) to in-plane (red and blue lines), at fixed $n = 0.8 \times 10^{12} cm^{-2}$ and $D = -0.4V/nm$. The data are only presented within the range with prominent oscillations, and are normalized to [0,1]. The curves are offset for clarity. **(b)** The ratio of $B_\theta$ and $B_0$ at different $\theta$, extracted from the FFT results from **(a)**, where $B_\theta$ is the FFT frequency at magnetic angle $\theta$, and $B_0$ is $B_\theta$ when magnetic field is completely out of plane. The blue line is the cosine curve. **(c)** $dV_y^{2\omega}/dB$ as a function of $1/B$ at four different $D$ with fixed carrier density. The black dashed lines are for denoting the consistent peak positions.

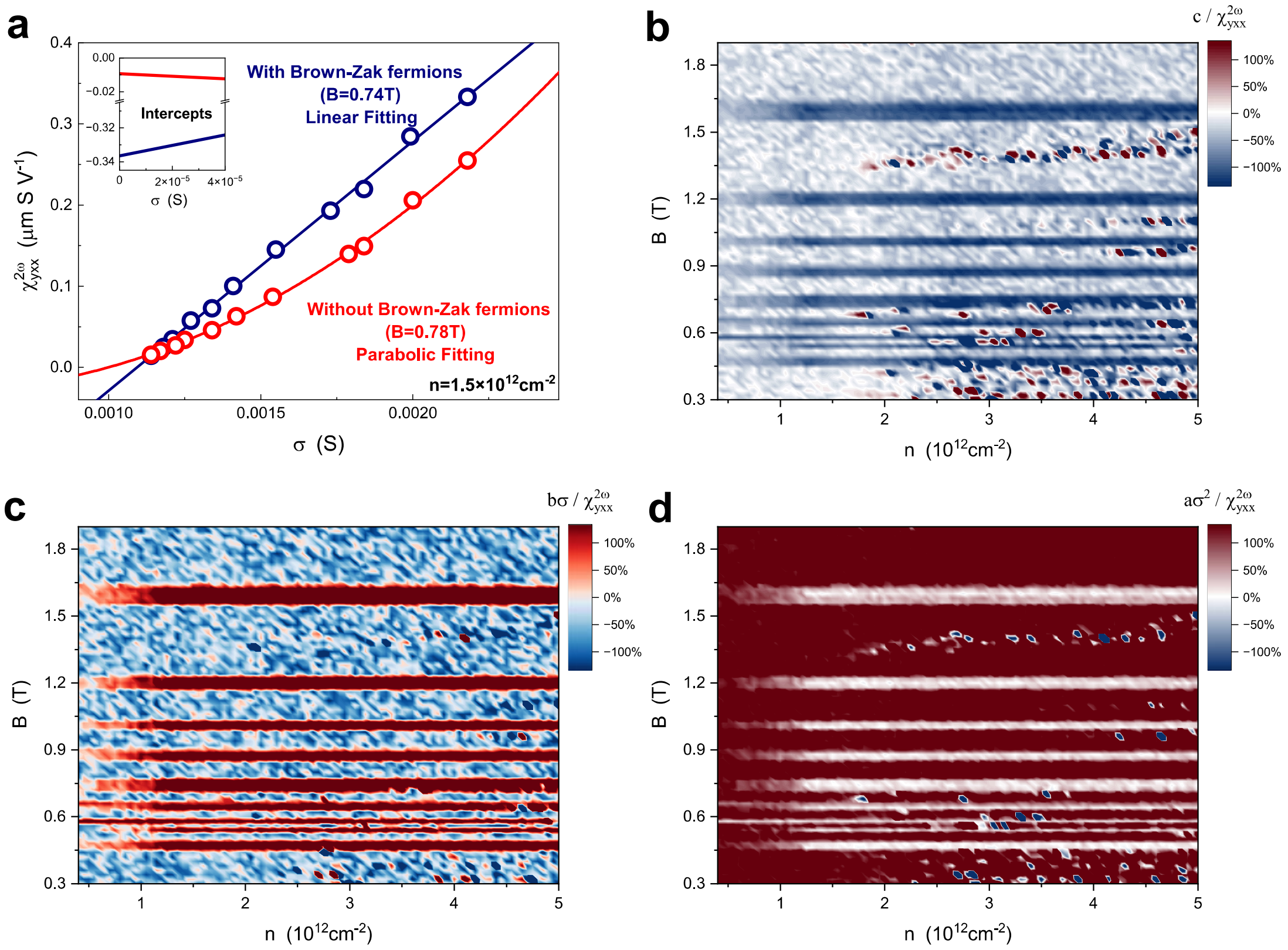

**FIG. 4. Scaling relation analysis of NHBZO in TDBG. (a)** The scaling relation analysis $\chi_{yxx}^{2\omega}\ vs\ \sigma$ at $n = 1.5 \times 10^{12} cm^{-2}$ and $D = -0.4V/nm$, by tuning temperature from $T = 1.6K$ to $10K$. The blue line and red curve are the linear fitting and parabolic fitting, respectively. Inset: the $\sigma \to 0$ regime of the fitting line and curve, showing the fitting intercepts $c$. **(b)-(d)** The $(n, B)$ map of scaling fitting coefficients $c, b$ and $a$. The scaling fittings are performed for fixed $(n, B)$ point via tuning temperature from $T = 1.6K$ to $10K$. We use the normalized quantity $c/\chi_{yxx}^{2\omega}$, $b\sigma/\chi_{yxx}^{2\omega}$ and $a\sigma^2/\chi_{yxx}^{2\omega}$ to show their relative proportion.